\documentclass[epjc3,5p,times,twocolumn,a4paper,showpacs,showkeys,amsmath,amssymb,superscriptaddress,nofootinbib]{svjour3} 

\usepackage{graphicx}
\usepackage{import}
\usepackage{txfonts}
\usepackage{epsfig}
\usepackage[numbers]{natbib}
\usepackage{listings}
\usepackage{mdwlist}
\usepackage{subfigure}
\usepackage[english]{babel}

\newcommand*{\megsign} {\mu^+ \to e^+ \gamma}

\newcommand*{\michelsign} {\mu^+ \to e^+ \nu\bar{\nu}}

\begin{document}

\title{Muon polarization in the MEG experiment: \\ predictions and measurements}

\subtitle{The MEG Collaboration}
\newcommand*{\INFNPi}{INFN Sezione di Pisa$^{a}$; Dipartimento di Fisica$^{b}$ dell'Universit\`a, Largo B.~Pontecorvo~3, 56127 Pisa; Scuola Normale Superiore$^{c}$, Piazza dei Cavalieri, 56127 Pisa Italy}
\newcommand*{\INFNGe}{INFN Sezione di Genova$^{a}$; Dipartimento di Fisica$^{b}$ dell'Universit\`a, Via Dodecaneso 33, 16146 Genova, Italy}
\newcommand*{\INFNPv}{INFN Sezione di Pavia$^{a}$; Dipartimento di Fisica$^{b}$ dell'Universit\`a, Via Bassi 6, 27100 Pavia, Italy}
\newcommand*{\INFNRm}{INFN Sezione di Roma$^{a}$; Dipartimento di Fisica$^{b}$ dell'Universit\`a ``Sapienza'', Piazzale A.~Moro, 00185 Roma, Italy}
\newcommand*{\INFNLe}{INFN Sezione di Lecce$^{a}$; Dipartimento di Matematica e Fisica$^{b}$ dell'Universit\`a del Salento, Via per Arnesano, 73100 Lecce, Italy}
\newcommand*{\ICEPP} {ICEPP, University of Tokyo 7-3-1 Hongo, Bunkyo-ku, Tokyo 113-0033, Japan }
\newcommand*{\UCI}   {University of California, Irvine, CA 92697, USA}
\newcommand*{\KEK}   {KEK, High Energy Accelerator Research Organization 1-1 Oho, Tsukuba, Ibaraki, 305-0801, Japan}
\newcommand*{\PSI}   {Paul Scherrer Institut PSI, CH-5232, Villigen, Switzerland}
\newcommand*{\Waseda}{Research Institute for Science and Engineering, Waseda~University, 3-4-1 Okubo, Shinjuku-ku, Tokyo 169-8555, Japan}
\newcommand*{\BINP}  {Budker Institute of Nuclear Physics of Siberian Branch of Russian Academy of Sciences, 630090, Novosibirsk, Russia}
\newcommand*{\JINR}  {Joint Institute for Nuclear Research, 141980, Dubna, Russia}
\newcommand*{\ETHZ}  {Swiss Federal Institute of Technology ETH, CH-8093 Z\" urich, Switzerland}
\newcommand*{\NOVST}  {Novosibirsk State Technical University, 630092, Novosibirsk, Russia}
\newcommand*{\NOVSU}  {Novosibirsk State University, 630090, Novosibirsk, Russia}

\date{Received: date / Accepted: date}

\author{A.~M.~Baldini~\thanksref{addr4}$^a$ \and
        Y.~Bao~\thanksref{addr1} \and
        E.~Baracchini~\thanksref{addr3}$^{\ddag}$ \and
        C.~Bemporad~\thanksref{addr4}$^{ab}$ \and
        F.~Berg~\thanksref{addr1,addr2} \and
        M.~Biasotti~\thanksref{addr5}$^{ab}$ \and
        G.~Boca~\thanksref{addr7}$^{ab}$ \and
        P.W.~Cattaneo~\thanksref{addr7}$^{a}$  \and
        G.~Cavoto~\thanksref{addr8}$^{a}$ \and
        F.~Cei~\thanksref{addr4}$^{ab*}$ \and
        G.~Chiarello~\thanksref{addr6}$^{ab}$ \and
        C.~Chiri~\thanksref{addr6}$^{a}$ \and
        A.~De Bari~\thanksref{addr7}$^{ab}$ \and 
        M.~De Gerone~\thanksref{addr5}$^{a}$ \and
        A.~D'Onofrio~\thanksref{addr4}$^{ab}$ \and
        S.~Dussoni~\thanksref{addr4}$^{a}$\and
        Y.~Fujii~\thanksref{addr3}  \and
        L.~Galli~\thanksref{addr4}$^{a}$ \and
        F.~Gatti~\thanksref{addr5}$^{ab}$ \and
        F.~Grancagnolo~\thanksref{addr6}$^{a}$ \and
        M.~Grassi~\thanksref{addr4}$^{a}$ \and
        A.~Graziosi~\thanksref{addr8}$^{ab}$ \and
        D.N.~Grigoriev~\thanksref{addr12,addr14,addr15} \and
        T.~Haruyama~\thanksref{addr9} \and
        M.~Hildebrandt~\thanksref{addr1} \and
        Z.~Hodge~\thanksref{addr1,addr2} \and
        K.~Ieki~\thanksref{addr1,addr3}  \and
        F.~Ignatov~\thanksref{addr12,addr15} \and
        T.~Iwamoto~\thanksref{addr3}  \and
        D.~Kaneko~\thanksref{addr3}  \and
        T.I.~Kang~\thanksref{addr11} \and
        P.-R.~Kettle~\thanksref{addr1} \and
        B.I.~Khazin~\thanksref{addr12,addr15}$^\dagger$ \and
        N.~Khomutov~\thanksref{addr13} \and
        A.~Korenchenko~\thanksref{addr13}  \and
        N.~Kravchuk~\thanksref{addr13}  \and
        G.M.A.~Lim~\thanksref{addr11} \and
        S.~Mihara~\thanksref{addr9}  \and
        W.~Molzon~\thanksref{addr11} \and
        Toshinori~Mori~\thanksref{addr3}  \and
        A.~Mtchedlishvili~\thanksref{addr1}  \and
        S.~Nakaura~\thanksref{addr3}  \and 
        D.~Nicol\`o~\thanksref{addr4}$^{ab}$ \and
        H.~Nishiguchi~\thanksref{addr9}  \and
        M.~Nishimura~\thanksref{addr3}  \and 
        S.~Ogawa~\thanksref{addr3}  \and
        W.~Ootani~\thanksref{addr3}  \and
        M.~Panareo~\thanksref{addr6}$^{ab}$ \and
        A.~Papa~\thanksref{addr1} \and
        A.~Pepino~\thanksref{addr6}$^{ab}$ \and
        G.~Piredda~\thanksref{addr8}$^{a}$ \and
        G.~Pizzigoni~\thanksref{addr5}$^{ab}$ \and
        A.~Popov~\thanksref{addr12,addr15} \and
        F.~Renga~\thanksref{addr8}$^{a}$ \and
        E.~Ripiccini~\thanksref{addr8}$^{ab}$ \and
        S.~Ritt~\thanksref{addr1} \and
        M.~Rossella~\thanksref{addr7}$^{a}$ \and
        G.~Rutar~\thanksref{addr1,addr2} \and
        R.~Sawada~\thanksref{addr3}  \and
        F.~Sergiampietri~\thanksref{addr4}$^{a}$ \and
        G.~Signorelli~\thanksref{addr4}$^{a}$ \and
        G.F.~Tassielli~\thanksref{addr6}$^{a}$ \and
        F.~Tenchini~\thanksref{addr4}$^{ab}$ \and
        Y.~Uchiyama~\thanksref{addr3} \and
        M.~Venturini~\thanksref{addr4}$^{ac}$ \and
        C.~Voena~\thanksref{addr8}$^{a}$ \and
        A.~Yamamoto~\thanksref{addr9} \and
        K.~Yoshida~\thanksref{addr3} \and
        Z.~You~\thanksref{addr11} \and
        Yu.V.~Yudin~\thanksref{addr12,addr15} \and
}

\institute{ \PSI \label{addr1} 
           \and
              \ETHZ \label{addr2}
           \and
              \ICEPP \label{addr3}
           \and
             \INFNPi \label{addr4}
           \and
             \UCI    \label{addr11}
           \and
             \INFNPv \label{addr7}
           \and
             \INFNRm \label{addr8}
           \and
             \INFNGe \label{addr5}
           \and
             \Waseda \label{addr10}
           \and
             \BINP   \label{addr12}
           \and
             \KEK    \label{addr9}
           \and
             \JINR   \label{addr13}
           \and
             \INFNLe \label{addr6}
           \and
             \NOVST  \label{addr14}
           \and
             \NOVSU  \label{addr15}
}

\thankstext[*]{e1}{Corresponding author: fabrizio.cei@pi.infn.it}
\thankstext[$\dagger$]{dagger}{Deceased.}
\thankstext[$\ddag$]{}{Presently at INFN, Laboratori Nazionali di Frascati, via E.Fermi, 40, 00044 Frascati (Roma) Italy}
\maketitle 

\begin{abstract}
The MEG experiment makes use of one of the world's most intense low energy 
muon beams, 
in order to search for the lepton flavour violating process 
$\mu^{+} \rightarrow {\rm e}^{+} \gamma$. 
We determined the residual beam polarization at the thin stopping target, 
by measuring the asymmetry of the angular distribution of Michel decay 
positrons as a function of energy. 
The initial muon beam polarization at the production is predicted to be 
$P_{\mu} = -1$ by the Standard Model (SM) with massless neutrinos. 
We estimated our residual muon polarization to be 
$P_{\mu} = -0.86 \pm 0.02 ~ {\rm (stat)} ~ { }^{+ 0.05}_{-0.06} ~ {\rm (syst)}$ 
at the stopping target, which is consistent with the SM predictions when the 
depolarizing effects occurring during the muon production, propagation and 
moderation in the target are taken into account.
The knowledge of beam polarization is of fundamental importance in order 
to model the background of our ${\megsign}$ search induced by the muon 
radiative decay: 
$\mu^{+} \rightarrow {\rm e}^{+} \bar{\nu}_{\mu} \nu_{\rm e} \gamma$.
\end{abstract}
\section{Introduction}\label{sec:intro}
Low energy muon physics experiments frequently use copious beams of {}\lq\lq 
surface muons\rq\rq, i.e. muons generated by pions decaying at rest close to 
the surface of the pion production 
target, such as those produced at meson factories (PSI and TRIUMF). 
In the Standard Model (SM) with massless neutrinos, positive (negative) muons 
are fully polarized, with the spin opposite (parallel) to the muon momentum 
vector, that is $P_{\mu} = -1$ for positive muons, at the production point; 
the muon polarization can be partially reduced by the muon interaction 
with the electric and magnetic fields of the muon beam line as well as 
with the muon stopping target. The degree of polarization at the muon decay 
point affects both the energy and angular distribution of the muon decay 
products i.e. Michel positrons and $\gamma's$ from the normal ${\michelsign}$ 
and radiative muon decay 
$\mu^{+} \rightarrow {\rm e}^{+} \bar{\nu}_{\mu} \nu_{\rm e} \gamma$. 
The muon decay products are an important background when searching for rare decays 
such as $\mu^{+} \rightarrow {\rm e}^{+} \gamma$; a precise knowledge of their 
distribution is therefore mandatory. We report on the determination of the 
residual muon polarization in the 
PSI $\pi$E5 \cite{PE5} channel and MEG beam line \cite{MEGdet} 
from the data collected by the MEG experiment 
between $2009$ and $2011$. Clear signs of the muon 
polarization are visible in the Michel positron angular distribution;  
the measured polarization is in good agreement with a theoretical 
calculation (see Section \ref{sec:polatheo}) based on the SM 
predictions and on the beam line characteristics. 

The MEG experiment at the Paul Scherrer Institute (PSI) \cite{PSI} has been 
searching for the lepton flavour violating decay 
$\mu^{+} \rightarrow {\rm e}^{+} \gamma$ since 2008. Preliminary 
results were published in \cite{meg2009,meg2010} and \cite{meg2013}. 
The analysis of the MEG full data sample is under way and will soon be published. 
A detailed description of the experiment can be found in 
\cite{MEGdet}. A high intensity surface muon beam 
($\sim 3\times 10^{7} \mu^+/{\rm s}$), from the $\pi$E5 channel 
and MEG beam line, is brought to rest in a $205~{\rm \mu m}$ 
slanted plastic target, placed at the centre of the experimental 
set-up. The muon decay products are detected by a spectrometer with a 
gradient magnetic field and by an electromagnetic calorimeter.
The magnetic field is generated by a multi-coil superconducting 
magnet (COBRA) \cite{thin-cable,ootani_2004}, with conventional compensation 
coils; the maximum intensity of the field is $1.26~{\rm T}$ at the 
target position. The positron momenta are measured by sixteen drift chambers 
(DCH) \cite{Hildebrandt2010111}, radially aligned, and their arrival 
times by means of a Timing Counter (TC) 
\cite{Dussoni2010387,DeGerone:2011te,DeGerone:2011zz},
consisting of two scintillator arrays, placed at opposite sides relative  
to the muon target. The momentum vector and the arrival time 
of photons are measured in a $900$ liter C-shaped liquid xenon photon 
detector (LXe) \cite{Sawada2010258,Mihara:2011zza}, equipped with a dense 
array of 846 UV-sensitive PMTs. A dedicated trigger system 
\cite{trigger2013,Galli:2014uga} allows an efficient preselection of possible 
$\mu^{+} \rightarrow {\rm e}^{+} \gamma$ candidates, with an almost 
zero dead-time. The signals coming from the DCH, TC and LXe detectors 
are processed by a custom-made waveform digitizer system (DRS4) 
\cite{ritt_2004_nim,Ritt2010486}
operating at a maximum sampling speed close to $2~{\rm GHz}$.
Several calibration tools are in operation, allowing a continuous monitoring 
of the experiment \cite{Baldini:2006,calibration_cw,papa_2010}. 
Dedicated prescaled trigger schemes collect calibration events 
for a limited amount of time (few hours/week). 
A complete list of the experimental resolutions ($\sigma$'s) for energies 
close to the kinematic limit $m_{\mu}/2$ can be found in \cite{meg2013}; the 
most relevant being: $\sim 340~{\rm keV}/c$ for the positron 
momentum, $\sim 10~{\rm mrad}$ for the positron zenith angle and 
$\sim 1$ and $\sim 3~{\rm mm}$ for the positron vertex along the two axes 
orthogonal to the beam direction. 

The beam axis defines the $z$-axis of 
the MEG reference frame. The part of the detector preceeding the muon target 
is called the {\it{\lq\lq UpStream\rq\rq}} (US) side and that 
following the muon target is called the {\it{\lq\lq DownStream\rq\rq}} 
(DS) side. The zenith angle $\theta$ of the apparatus ranges from 
$\approx 60^{\circ}$ to $\approx 120^{\circ}$, with $\left( 60^{\circ} - 
90^{\circ}\right)$ defining the DS-side and $\left( 90^{\circ} - 
120^{\circ}\right)$ defining the US-side. The SM prediction is 
$P^{z}_{\mu} = -1$ for muons travelling along the positive $z$-axis. 
\section{Theoretical issues}\label{sec:polatheo}
The $\pi$E5 channel is a high-intensity low-energy pion and muon beam line 
in the $10~{\rm MeV}/c < p < 120~~{\rm MeV}/c $ momentum range. Surface muons 
have a kinetic energy of $4.12~{\rm MeV}$ and a muon momentum of 
$\approx 29.79~{\rm MeV}/c$ and are produced fully polarized along the 
direction opposite to their momentum vector. Several depolarizing effects 
can reduce the effective polarization along the beam line. They are 
classified into three groups:
\begin{itemize}
\item [1)] effects at the production stage, close to and within the 
production target; 
\item [2)] effects along the beam line up to the stopping target; 
\item [3)] effects during the muon moderation and stopping process in the target. 
\end{itemize}
\subsection{Depolarization at the production stage}
Since the angular divergence of the beam is not zero, the average 
muon polarization $P_{\mu}$ along the muon flight direction does not coincide with 
$P^{z}_{\mu}$ where $z$ is the direction of the muon beam 
(the beam acceptance at the source is $150~{\rm msr}$ and the angular divergence 
is $450~{\rm mrad}$ in the horizontal and $120~{\rm mrad}$ in the vertical 
direction). 

One such depolarizing effect is due to the multiple scattering in the target, 
which modifies the muon direction leaving the spin unaffected.  
Surface muons have a maximum range in the carbon production target of 
$0.82~{\rm mm}$. The average broadening angle due to multiple scattering 
is then given by (see for instance \cite{Pifer}):
\begin{equation}
\langle \cos \alpha \rangle = 1 - 21\frac{l}{X_{0}} \left[ \left( 
\frac{30}{P}\right)^{0.5} - 1\right],
\label{multiscat}
\end{equation}
where $P$ is the muon momentum in ${\rm MeV}/c$ and $l/X_{0}$ is the muon 
path in the target in units of carbon radiation lengths ($X_{0} = 18.8~{\rm cm}$).
We obtain $\langle \cos \alpha \rangle = 0.997$, a contribution of less 
than $0.5~\%$. 

A more important effect is due to {\it{\lq\lq cloud muons\rq\rq}}, i.e. 
muons originating from pion decays in flight, in or close to the production 
target, and accepted by the beam transport system. These muons have only 
a small net polarization due to their differing acceptance kinematics 
which leads to an overall reduction of the beam polarization, based on studies 
performed at LAMPF \cite{VanDyck:1979xr} and measurements we made at the $\pi$E5 channel 
at PSI. The latter involved the fitting of a constant cloud muon content 
to the limited region of the measured muon momentum spectrum, around the 
kinematic edge at $\approx 29.79~{\rm MeV}/c$. This was cross-checked by 
direct measurements of negative cloud muons at the MEG central beam 
momentum of $28~{\rm MeV}/c$, where there is no surface muon contribution 
on account of the charge sign (muonic atom formation of stopped negative 
muons). The cloud muon content was found to be consistent from both measurements 
when taking the kinematics and cross-sections of positive and negative pions 
into account. This leads to an estimated depolarization of 
$\left( 4.5 \pm 1.5~\% \right)$, which is the single-most important 
effect at the production stage. 
\subsection{Depolarization along the beam line}
The MEG beam line comprises of several different elements: quadrupole and 
bending magnets, fringing fields, an electrostatic separator, a 
beam transport solenoid and the COBRA spectrometer. The equation of 
motion of the muon spin $\vec{s}$ is described, even in a spatially 
varying magnetic field such as the COBRA spectrometer, 
by the Thomas equation \cite{Jackson}:
\begin{eqnarray}
\lefteqn{\frac{d\vec{s}}{dt} = \frac{e}{mc} \vec{s} \times 
\left[ \left( \frac{g}{2} - 1 + \frac{1}{\gamma} \right)\vec{B} - 
\left( \frac{g}{2} - 1 \right) \frac{\gamma}{\gamma + 1} 
\left( \vec{\beta} \cdot \vec{B} \right) \vec{\beta} \right . } \hspace{1cm}\nonumber\\
& & \left. - \left( \frac{g}{2} - \frac{\gamma}{\gamma + 1}\right) 
\vec{\beta} \times \vec{E} \right],
\label{thomas}
\end{eqnarray}
where $\beta$, $e$ and $m$ are the muon velocity, electric charge and 
mass, $c$ is the speed of light, $\gamma = 
1/\sqrt{\left( 1 - \beta^{2} \right)}$, $g$ is the muon gyromagnetic factor 
and $\vec{B}$ and $\vec{E}$ are the electric and magnetic field vectors. 
In principle this equation is valid only for uniform fields, but 
it gives correct results even in our case since any effect due to 
the non-uniformity of the magnetic field is many orders of magnitude 
smaller than the Lorentz force in the weak gradient field of 
COBRA. 
From Eq.~\ref{thomas} we can obtain the time evolution of the longitudinal 
polarization, defined as the projection of the spin vector along the 
momentum vector, which is given by:
%
%
\begin{equation}
\frac{d\left( \vec{s} \cdot \vec{\beta} \right)}{dt} = - \frac{e}{mc} \vec{s}_{\perp} \cdot 
\left[ \left( \frac{g}{2} - 1 \right)\vec{\beta} \times \vec{B} + 
\left( \frac{g \beta}{2} - \frac{1}{\beta}\right) \vec{E} \right],
\label{thomaslong}
\end{equation}
where $\vec{s}_{\perp}$ is the projection of the spin vector in the plane 
orthogonal to the muon momentum. 
In this equation, the first 
contribution is due to the muon magnetic moment anomaly 
($\frac{g - 2}{2} \approx \frac{\alpha}{2 \pi}$) and the second to 
the presence of an electric field. 
In the MEG beam line the first term is associated with the guiding elements  
(quadrupole and bending magnets), while the second term 
is associated with the electrostatic separator. The geometrical parameters of the 
beam elements and their field intensities are \cite{PE5}: 
for the deflecting magnets the length is $\approx 70~{\rm cm}$ and the 
vertical field is $\approx 0.15~{\rm T}$; for the electrostatic separator 
the length is $82~{\rm cm}$, the gap between the plates $19~{\rm cm}$ and 
the applied voltage $-195~{\rm kV}$. The COBRA spectrometer has a weak 
spatially varying magnetic field, which muons are subjected to while 
travelling on the US-side of the magnet, after being 
focused by the beam transport solenoid; the average vertical 
component of the COBRA magnetic field around the muon trajectory is of order 
of $0.025~{\rm T}$ and its contribution to the spin rotation is about 
one order of magnitude smaller than the one of the bending magnets. 
With these parameters we evaluated a spin rotation of $\approx 0.25^{\circ}$ 
due to the magnetic component and of $\approx 7^{\circ}$ due to the 
electrostatic component. Note that the longitudinal polarization is, 
by definition, referred to the muon velocity, while the polarization 
we are interested in is the one in the beam direction, our natural 
quantization axis. Therefore the spin rotation results in a depolarizing 
effect of $\approx 0.8~\%$; this is confirmed by a numerical integration 
of the Thomas and Lorentz equations along the MEG beam line. 
\subsection{Depolarization during the muon moderation and stopping processes.}
The largest muon depolarization effect is expected to take place 
in the MEG muon stopping target. The behaviour of positive 
muons in matter is extensively discussed in the literature (for a 
review \cite{Brewer}). After a rapid moderation and thermalization 
of muons in matter, muonium ($\mu^{+} {\rm e^{-}}$) is formed and 
further thermalized by collisions. The muon polarization is unaffected 
during the muonium formation and thermalization and subsequent decay.  
Muonium interaction with the magnetic field in vacuum is 
described by a hyperfine Hamiltonian, which includes the muon-electron 
spin-spin interaction and the Larmor interaction of both spins with the external 
field. On the basis defined by the total spin $S$ and by its projection 
along the quantization axis $S_{Z}$, the muonium wavefunction is a 
superposition of a triplet state ($S = 1$) and of a singlet state ($S = 0$). 
If one assumes muons to be fully polarized in the longitudinal 
direction when they enter the target and electrons in the target 
to be unpolarized, the initial state of the muonium formation is a 
$50~~\%-50~~\%$ mixture of the state $\left( S = 1, S_{Z} = -1 \right)$ and 
the combination of $\left( S = 1, S_{Z} = 0\right)$ and 
$\left( S = 0, S_{Z} = 0\right)$. The coefficients of this combination 
and their time evolution can be calculated as functions of the 
ratio $x = B/B_{0}$, where $B$ is the external magnetic field and 
$B_{0} = 0.1585~~{\rm T}$. While the $\left( S = 1, S_{Z} = -1 \right)$ 
component is a pure state and is constant, the other oscillates 
with time; one can calculate its time average, 
which translates into an average longitudinal polarization given by:
\begin{equation}
\langle P_{\parallel} \left( x \right) \rangle = \frac{1}{2} 
\left(1 + \frac{x^{2}}{1 + x^{2}} \right). 
\label{avepol}
\end{equation}
Since $x$ at the position of the MEG target is $x \approx 7.9$, we 
obtain an average residual polarization of $99.2~\%$: any depolarizing 
effect is quenched by the strong magnetic field. 
However, muons are propagating in a dense medium and not in vacuum; 
therefore the muonium interaction with the material medium should be 
taken into account, making a detailed calculation impossible. 
We therefore used available experimental data, 
i.e. direct measurements of the muon residual polarization after crossing 
different targets immersed in external magnetic fields. The MEG target is 
a layered structure of polyethylene and polyethylene terephthalate (PET), 
for which no direct measurement is available; we assume this 
material to behave like polyethylene \cite{Swanson,Buhler1,Buhler2}. 

\begin{figure}[htb]
\includegraphics[width=0.5\textwidth]{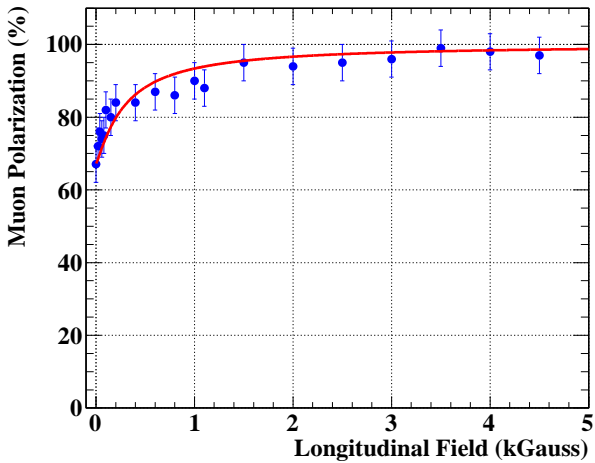}\hspace{2pc}
\caption{\label{fig:residualpol} Muon residual polarization after the muons 
stop in a polyethylene target, as a function of the external magnetic field 
(adapted from \cite{Buhler2}).}
\end{figure}

With zero magnetic field, the residual muon polarization 
is $\left( 67.1 \pm 2.0 \right)~\%$ and reaches $\approx 100~\%$ 
for increasing magnetic fields. Figure~\ref{fig:residualpol} shows 
the value of muon residual polarization as a function of the magnetic field 
intensity (adapted from \cite{Buhler2}): the polarization saturates 
at $\approx 100~\%$ for a magnetic field intensity of $\approx 4~{\rm kG}$, 
while the central value of the COBRA magnetic field is $> 12~{\rm kG}$. 
So, we can assume that even in our case the strong magnetic field quenches 
any depolarizing effect. 

The last point to be addressed is that muons reach the target centre under 
different angles within a $\approx 1 \times 1~{\rm cm}^{2}$ beam spot. This 
angular spread corresponds to an apparent depolarization, since $P_{\mu}$ 
does not coincide with $P^{z}_{\mu}$. Using the 
full MEG Monte Carlo (MC) simulation we evaluated that the angular divergence at the 
target corresponds to a cone of $< 20^{\circ}$ opening angle, corresponding to 
$\approx 3~\%$ apparent depolarization. 
\subsection{Total depolarization}
In conclusion, the main depolarizing effects are due to cloud muons 
and beam divergence. The average final polarization along the beam axis ($z$) is:
\begin{equation}
\langle P^{z}_{\mu} \rangle = \left( - 0.91 \pm 0.03 \right),
\label{polaval}
\end{equation}
where the systematic uncertainty takes into account the uncertainties 
in this computation. The various contributions are listed in 
Tab.~\ref{tab:depolarization}.
\begin{table}
\caption{\label{tab:depolarization} Summary of main depolarizing effects (\%).}
\begin{tabular*}{\columnwidth}{@{\extracolsep{\fill}}lc@{}}
\hline\hline
Source & (\%) \\\hline
Multiple scattering in the production target & 0.3\\
Cloud muons & 4.5\\
Muon transport along the beam line & 0.8\\
Muon interactions with the MEG target & negligible \\
Muon angular spread at the target & 3.0\\
\hline
Total & 8.6\\
\hline\hline
\end{tabular*}
\end{table}
\section{Expected Michel positron spectrum from polarized muons}\label{sec:angdist}
\begin{figure}[hbt]
\includegraphics[width=0.5\textwidth]{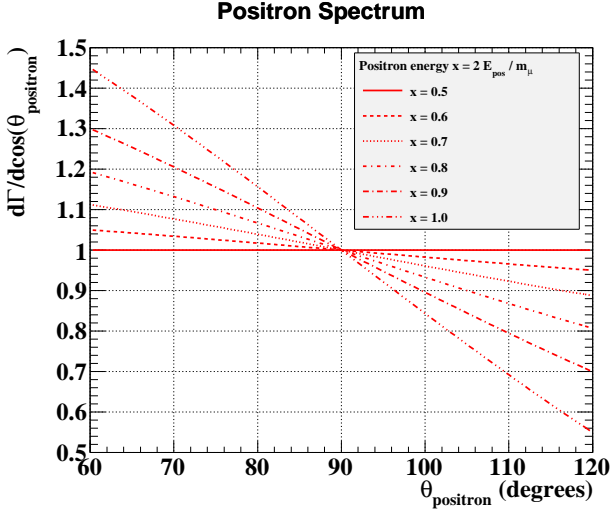}\hspace{2pc}
\caption{\label{fig:micheldis} Angular distributions of Michel positrons referred 
to the $\mu^+$ spin direction for six 
different values of $x$, as given by Eq.~\ref{micheldist}. The vertical 
scale is in arbitrary units and normalized to $1$ for 
$\theta_{\rm e} = 90^{\circ}$.}
\end{figure}

The angular distribution of Michel positrons was calculated in detail by several 
authors including the effect of the electron mass and the first order radiative corrections 
\cite{kuno_2001, kinoshita_1959, Arbuzov:2001ui}. The bidimensional energy-angular distribution for polarized $\mu^+$ 
decaying at rest, neglecting the electron mass, takes the following form:
\begin{eqnarray}
\lefteqn{\frac{d^{2}\Gamma \left( \mu^+ \rightarrow {\rm e}^{+} \nu \bar{\nu} \right)}
{dx d \cos \theta_{\rm e}} =
\frac{{m_{\mu}}^{5} {G_{F}}^{2}}{192 {\pi}^{3}} }
\nonumber\\
& & x^{2} \left[ F(x) 
+ P_{\mu} \cos \theta_{\rm e} G(x) \right] 
\nonumber\\
F(x) & & = f_0(x) + \frac{\alpha}{2\pi} f_1(x) + {\cal O}(\alpha^2) \nonumber\\
G(x) & & = g_0(x) + \frac{\alpha}{2\pi} g_1(x) + {\cal O}(\alpha^2) \nonumber\\
f_0(x) &= & \left( 3 - 2 x \right) \qquad g_0(x) = \left( 2 x - 1 \right) 
\label{micheldist}
\end{eqnarray}
where $P_{\mu}$ is 
the $\mu^+$ polarization along a selected axis, $x = 2E_{{\rm e}^{+}}/m_{\mu}$ ($0 \le x \le 1$) 
and $\theta_{\rm e}$ is the angle formed by the positron momentum vector and the 
polarization axis. Expressions for $f_1(x)$ and $g_1(x)$ neglecting the electron mass are
available in \cite{kinoshita_1959}; the MC simulations in the following are based on Eq.~\ref{micheldist}
including first order radiative corrections and neglecting the electron mass. 
Formulae incorporating the dependence on electron 
mass for all terms in Eq.~\ref{micheldist} are presented in \cite{Arbuzov:2001ui}.\\ 
We show in Fig.~\ref{fig:micheldis} the angular 
distribution from Eq.~\ref{micheldist} in the range $60^{\circ} \div 120^{\circ}$ for different values of $x$. 
The differential decay width for $x = 1$ at $\theta_{\rm e} = 70^{\circ}$ is 
about twice that at $\theta_{\rm e} = 110^{\circ}$. Inspection of 
Fig.~\ref{fig:micheldis} shows that detectable effects are expected in the 
MEG data sample, even if the MEG apparatus is not the best suited for 
polarization measurements due to the relatively small angular range, 
centred around $\theta_{\rm e} = 90^{\circ}$. 
\section{Results of the measurement}
\subsection{Generalities}
In the previous section we showed that polarization effects can be observed 
in the angular distributions of high-energy positrons from Michel decays. 
In addition to that, 
the distribution of high-energy photons from Radiative Muon Decay (RMD) 
is expected to be affected by the polarization; 
however its 
associated error is very large, because of the intrinsic uncertainties in the 
analysis method, mainly related to the determination of the photon emission 
angle, and because of the presence in this data sample of a large background 
of photons from other sources (e.g. bremsstrahlung, annihilation in flight, 
pile-up of lower energy gamma's ...). 
We will therefore disregard this item. 

It is important to note that in Eq.~\ref{micheldist} the quantization 
axis is the muon spin direction; however, surface muons are expected 
to be fully polarized in the backward direction, i.e. along the negative 
$z$-axis. Therefore, the polar angle $\theta$ in the MEG reference frame 
is related to $\theta_{\rm e}$ in Eq.~\ref{micheldist} by 
$\theta = 180^{\circ} - \theta_{\rm e}$. Hence, the excess in the theoretical 
angular distribution Eq.~\ref{micheldist} for $\theta_{\rm e} < 90^{\circ}$ 
corresponds to an excess for $\theta > 90^{\circ}$ in the experimental angular 
distribution, i.e. on the US-side. 

\begin{figure}[hbt]
\includegraphics[width=0.5\textwidth]{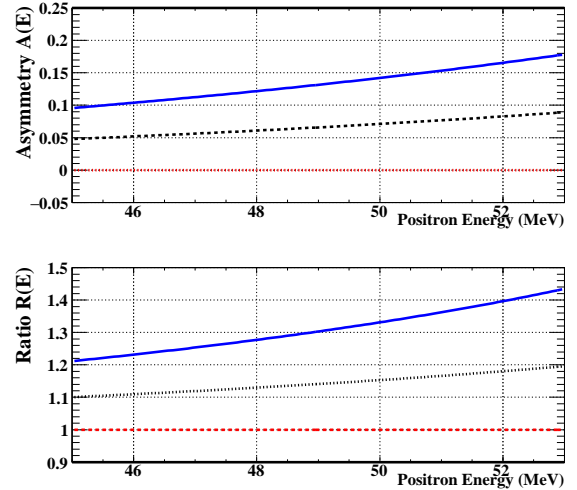}\hspace{1pc}
\caption{\label{fig:michelasyrat} 
The asymmetry (\ref{asyformula}) 
(upper plot) and the ratio (\ref{ratformula}) (lower plot) as a function 
of the positron energy $E_{\rm e^{+}}$, for three representative 
polarization values: $0$ (red dotted line), $-0.5$ (black dashed line) 
and $-1$ (blue continuous line). All spectra are obtained by 
integrating the 2-D (energy, angle) distributions over the MEG angular range. 
The contribution of first order R.C. on both asymmetry and ratio is $\approx 0.3\,\%$ 
while those of higher order corrections and of neglecting the electron 
mass are even smaller.} 
\end{figure}

A very powerful way to study the muon polarization is to compare the 
energy spectra, integrated over the angular acceptance, on the US 
($\left(dN/dE_{\rm e^{+}}\right)_{US}$) and on 
the DS ($\left(dN/dE_{\rm e^{+}}\right)_{DS}$) sides. 
In Fig.~\ref{fig:michelasyrat} we show the expected asymmetry between $45~{\rm MeV}$ and $53~{\rm MeV}$
as a function of positron energy $E_{\rm e^{+}}$   :
\begin{equation} 
A\left( E_{\rm e^{+}} \right) = \frac
{\left( 
\left(dN/dE_{\rm e^{+}}\right)_{US} - \left(dN/dE_{\rm e^{+}}\right)_{DS} \right)}
{\left( 
\left(dN/dE_{\rm e^{+}}\right)_{US} + \left(dN/dE_{\rm e^{+}}\right)_{DS} \right)}
\label{asyformula}
\end{equation}
in the upper part and the ratio:
\begin{equation}  
R\left( E_{\rm e^{+}} \right) = 
\frac
{\left(dN/dE_{\rm e^{+}}\right)_{US}}
{\left(dN/dE_{\rm e^{+}}\right)_{DS}}
\label{ratformula}
\end{equation}
in the lower part for three representative polarization values: $0$ 
(red dotted line), $-0.5$ (black dashed line) and $-1$ 
(blue continuous line). First order R.C. are taken into account 
and have a $\approx 0.3\,\%$ effect on both asymmetry and ratio. 
\subsection{Analysis of Michel positrons}\label{michpol}
\label{sect:analysis}
Experimentally measured angular distributions are a result of the convolution 
of the expected theoretical distributions with the detector response, 
acceptance and thresholds, whose non-uniformities can mimic angular 
asymmetries or create fictitious ones. 
To\-po\-logical requirements and quality cuts needed to define 
and fit charged particle tracks also introduce angle-dependent 
non-uniformities. In particular, the tracking algorithm has a lower 
efficiency for positrons emitted with small longitudinal momenta, resulting 
in a dip in the angular distribution of Michel positrons for $\theta \approx 
90^{\circ}$ (see later Fig.~\ref{fig:compathetadouble}). MEG positrons are mainly 
produced by muon decays in the stopping target, with a significant fraction 
($\sim 20~\%$) decaying off-target, in beam elements or in the surrounding  helium gas. 
However, this contribution can be 
minimized by requiring the reconstructed positron decay vertex to lie within 
the target volume. The fraction of the positrons decaying off-target and 
reconstructed on the target was evaluated by a complete MC simulation of the 
muon trajectory along the PSI/MEG beam line up to the stopping target and 
of the subsequent muon decay. 
This fraction was found to be smaller than 0.5\%\ and can be considered as a source of 
systematic uncertainty assuming, very conservatively, the same effect on the polarization 
measurement. 
In summary, an analytical prediction of the 
experimental distribution is rather complicated; hence, we decided to 
measure the muon polarization by 
means of two different analysis strategies:
\begin{itemize}
\item in the first one, we compared the energy integrated experimental angular 
distribution of Michel positrons with that obtained by a detailed Geant3-based 
MC simulation of those events, as seen in the MEG detector, with the muon 
polarization as a free input parameter;
\item in the second one, we measured the US-DS asymmetry 
$A\left( E_{\rm e^{+}} \right)$ and the ratio $R\left( E_{\rm e^{+}} \right)$ 
as a function of positron energy and fit them with the expected 
phenomenological forms, after unfolding the detector 
acceptance and response. 
\end{itemize}
\subsubsection{MC simulation}
The MEG MC simulation is described in details in \cite{meg2009} and 
\cite{softwaretns}. Michel positrons were generated in the stopping target 
(the full simulation of the muon beam up to the stopping target described above 
was not used since it is much slower and does not bring significant advantages 
in this case) with a minimum energy of $40~{\rm MeV}$ and a muon 
polarization $P_{\mu}$ varying between $0$ and $-1$ in steps of $0.1$. 
A smaller step size of $0.05$ was used between $-0.8$ and $-1$, close 
to the expected value (section \ref{sec:polatheo}). Separate samples of 
MC events were produced for each polarization value and the 
positron energy and direction were generated according to the theoretical 
energy-angle distribution corresponding to this polarization. Positrons 
were individually followed within the fiducial volume and their hits in the 
tracking system and on the timing counters were recorded; 
a simulation of the electronic chain converted these hits into anodic 
and cathodic signals which were processed by the same analysis 
algorithms used for real data. 
Modifications of the apparatus configuration during the whole period of 
data taking were simulated in detail, following the information recorded 
for each run in the experiment database. The position and spatial orientation 
of the target varied slightly each year, as well as trigger and acquisition 
thresholds, beam spot centre and size and the drift chamber alignment 
calibration constants. Some of the drift chambers suffered from 
instabilities, with a time scale from days to 
weeks, with their supply voltages finally set to a value 
smaller than nominal. The supply voltage variations, chamber by chamber, 
were also followed in the simulation on a run by run basis. However, 
voltage instabilities do not significantly affect the polarization measurement. 
Since drift chamber wires run along the $z$-axis, a non operating chamber produces 
the same effect on US and DS if the beam is perfectly centred on the target, 
while it gives a second order contribution to the US-DS asymmetry when the 
beam is not perfectly centred. The number of MC events generated using 
the global configuration (target position, alignment ...) corresponding 
to a given year is proportional to the actual amount of data 
collected in that year. 
\subsubsection{Data sample}
The data sample contains the events collected between $2009$ and $2011$ by a 
pre-scaled trigger requiring only a timing counter hit above the threshold 
(so called {\it{\lq\lq trigger 22\rq\rq}}). The analysis procedure requires 
an accurate pre-selection of good quality tracks: strict selection cuts 
are applied in order to single out tracks with good angular and momentum 
resolutions, 
well matched with at least one timing counter hit and with the decay vertex 
reconstructed within the target volume. A fiducial volume cut is 
included to avoid efficiency distorsions at the borders of the acceptance. 
The sample and the selection criteria are essentially those used to identify 
Michel events for the absolute normalization of the MEG data (see 
\cite{meg2010,meg2013}). About $37k$ ($2009$), $65k$ ($2010$) and $115k$ ($2011$) 
positron tracks passed all selection cuts, for a total of about 
$2.1 \times 10^{5}$ events. The same criteria were applied to the MC 
tracks; about $1.3 \times 10^{5}$ events passed all selections 
for each polarization value. 
\subsubsection{Comparison between MC and data}
\begin{figure}[hbt]
\includegraphics[width=0.5\textwidth]{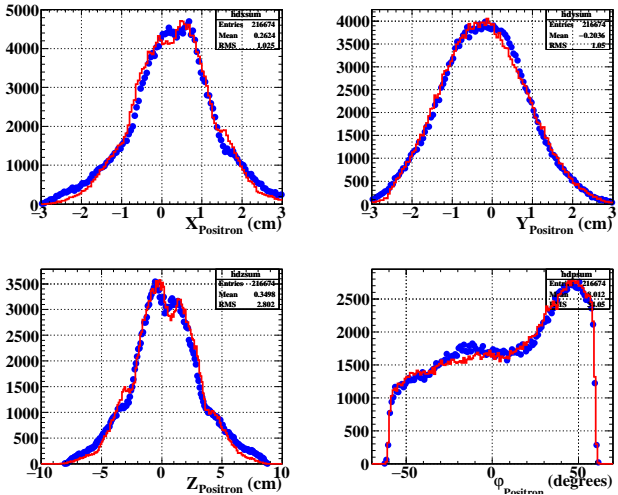}
\caption{\label{fig:datamcxyzphi} Comparison between the reconstructed vertex 
coordinates $x$, $y$ and $z$ and the azimuthal $\phi$ angle, for real 
(blue points) and simulated data (red line, normalized to the data). 
The polarization of the MC sample is $-0.85$; the distributions corresponding 
to different polarizations are almost identical.} 
\end{figure}

\begin{figure}[hbt]
\includegraphics[width=0.5\textwidth]{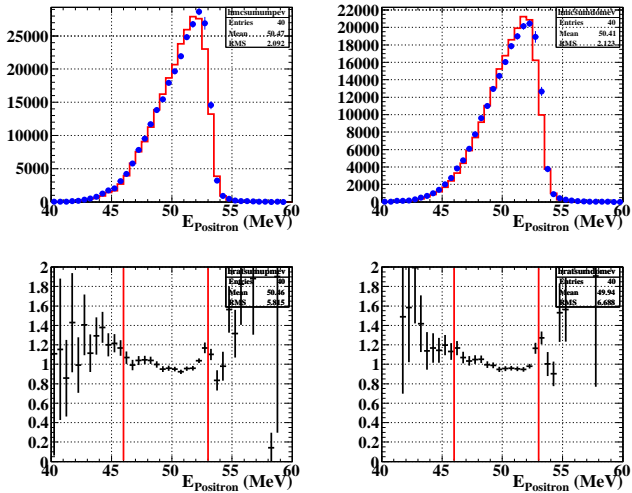}
\caption{\label{fig:datamcposene} Upper plots: comparison between data 
(blue points) and MC (red line, normalized to the data) positron energy 
spectra for the US (left) and DS (right) sides. Lower plots: ratio 
data/MC as a function of positron energy for US (left) and DS (right) 
sides. The red vertical lines indicate the boundaries of the fitting region 
($46~{\rm MeV} < E < 53~{\rm MeV}$)}.
\end{figure}

The comparisons between the reconstructed positron vertex coordinates 
$x$, $y$ and $z$ for data (blue points) and MC (red line, normalized to the 
data) are shown in Fig.~\ref{fig:datamcxyzphi}, top and bottom left; 
at the bottom right the same comparison 
for the reconstructed azimuthal angle $\phi$ at the positron emission 
point is shown. 
We also show in Fig.~\ref{fig:datamcposene} the comparison between data 
(blue points) and MC (red line) positron energy spectra on the US (left) 
and DS (right) sides.
In the upper part of the figure we report the superimposed data and MC 
distributions, while in the lower part we show the ratios data/MC as a 
function of the positron energy (in ${\rm MeV}$). All spectra are corrected 
for the left-right correction factors which will be discussed in the next
section.  
The red vertical lines in the bottom plots define the energy region 
where the polarization fit is performed ($46~{\rm MeV} < E < 53~{\rm MeV}$). 
The agreement between data and MC is generally quite good for the spatial 
coordinates, while some ($< 10\,\%$) discrepancies can be observed in the 
energy spectra and expecially in their ratios, even in the fit region. 
Data/MC ratios are consistent with unity for $48~{\rm MeV} < E < 52~{\rm MeV}$, 
but exhibit some systematic differences close to the threshold ($E \approx 
45-48~{\rm MeV}$) and in the upper edge ($E > 52~{\rm MeV}$). Such discrepancies 
are due to the fact that the MC simulation is not able to perfectly reproduce 
the experimental energy resolution: for instance $\sigma_{E} \approx 
340~{\rm keV}$ for data and $\approx 260~{\rm keV}$ for MC at 
$E = 52.83~{\rm MeV}$. However, if one looks at both bottom plots together, 
one sees that the differences are clearly correlated; then, they tend to 
cancel out when one uses $A\left( E_{\rm e^{+}} \right)$ or  
$R\left( E_{\rm e^{+}} \right)$ as analysis tools. 
We also note that the differences are particularly relevant in the year $2010$ 
sample, when the beam centre was displaced with respect to the target centre 
by some ${\rm mm}$. (See section \ref{sec:syst} dedicated 
to the analysis of systematic uncertainties.)  

The general agreement between data and MC for all reconstructed variables 
demonstrates our ability to correctly simulate the behaviour of the apparatus. 
\subsubsection{Efficiency correction for MC and data}
The efficiency for the full reconstruction of a positron event 
is composed of two parts: the absolute efficiency $\epsilon(Track)$ 
for producing a track satisfying all trigger and software requirements 
and the relative 
efficiency $\epsilon \left( TC \left| Track \right. \right)$ of having 
a TC hit, given a track. Both efficiencies are functions of the 
positron energy and emission angles and can be different on the US and DS 
sides because of intrinsic asymmetries 
of the experimental apparatus. 

The $\epsilon \left( TC \left| Track \right. \right)$ efficiency was 
separately computed for MC and real events. In the case of MC this 
calculation is straightforward. In the more complicate case of real 
data,  we selected positrons collected by a different pre-scaled trigger 
(so called {\it{\lq\lq trigger 18\rq\rq}}) requiring only loose conditions on 
the number and the topological sequence of fired drift chambers, 
and selected the fraction of tracks with an associated good TC hit 
within this sample. The MC and data $\epsilon \left( TC \left| Track 
\right. \right)$ efficiency matrices were then used to correct the 
$\theta$ angular distributions, $A\left( E_{\rm e^{+}} \right)$ 
and $R\left( E_{\rm e^{+}} \right)$. 

\begin{figure}[hbt]
\includegraphics[width=0.5\textwidth]{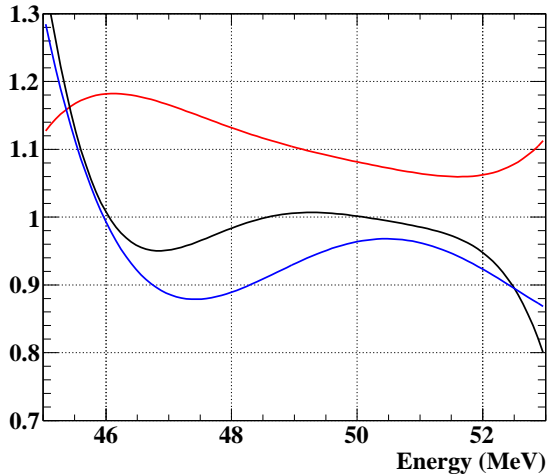}
\caption{\label{fig:corrfunc} Empirical correction functions 
applied to make the ratio $R\left( E_{\rm e^{+}} \right)$ 
for MC events generated with null polarization consistent with 
unity. The red line is for $2009$, the black one for $2010$ 
and the blue line for $2011$ sample.}
\end{figure}

The $\epsilon(Track)$  efficiency was extracted from MC by looking at the 
reconstructed $R\left( E \right)$ in the MC sample generated with 
$P_{\mu} = 0$ and determining, year by year, an empirical correction 
function which makes this $R\left( E \right)$ always consistent with 
unity within the errors. We show in Fig.~\ref{fig:corrfunc} the correction 
functions for $2009$ (red), $2010$ (black) and $2011$ (blue) samples. 
We then applied the same correction functions to all MC samples and 
we checked that the polarization values extracted by fitting 
$A\left( E_{\rm e^{+}} \right)$ and $R\left( E_{\rm e^{+}} \right)$ were 
consistent with those generated. 
The correction functions were also applied to the data since the good agreement 
between MC and data shown in Fig.~\ref{fig:datamcxyzphi} and 
\ref{fig:datamcposene} gives us confidence of the correct apparatus response 
to positron events. 
\subsubsection{Results of first strategy: angular distribution}
\begin{figure}[hbt]
\includegraphics[width=0.5\textwidth]{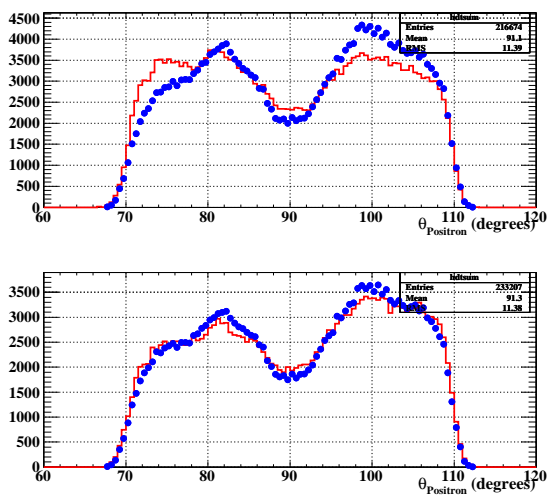}
\caption{\label{fig:compathetadouble} Comparison between the angular 
distributions for simulated (red line) and measured (blue points) Michel 
positrons; $P_{\mu} = 0$ (upper plot) and $P_{\mu} = -0.85$ (lower plot). 
The polar angle $\theta$ is referred to the beam axis. Histograms are 
normalized to the data.}
\end{figure}

In Fig.~\ref{fig:compathetadouble} the comparison between the angular 
distributions of real data (blue points) and of MC events (red line, 
normalized to the data), after inserting the matching efficiency 
corrections, as a function of $\theta$ angle for two different 
polarization values is shown: $P_{\mu} = 0$ in the upper plot 
and $P_{\mu} = -0.85$ in the lower plot. 
According to Eq.~\ref{micheldist} and to the definition of $\theta$, we 
expect to observe an asymmetric distribution for large values of the 
polarization, with an excess on the US-side ($\theta > 90^{\circ}$) and a 
symmetric distribution for null polarization. 
Fig.~\ref{fig:compathetadouble} shows a clear disagreement between data and MC 
for $P_{\mu} = 0$ and a good agreement for $P_{\mu} = -0.85$. The simulation 
well reproduces the US-DS asymmetry observed in the data, as well as the 
dip for $\theta \approx 90^{\circ}$. Angular distributions for 
$P_{\mu} = -0.8$ and for 
$P_{\mu} = -0.9$ do not significantly differ from that shown for 
$P_{\mu} = -0.85$: the comparison between data and 
MC gives strong indications for a large polarization, 
$- \left(0.8 - 0.9 \right)$, but it is not precise enough to 
single out a value of $P_{\mu}$, with its uncertainty.
\subsubsection{Results of second strategy: US-DS asymmetry and ratio}
\begin{figure}[hbt]
\includegraphics[width=0.5\textwidth]{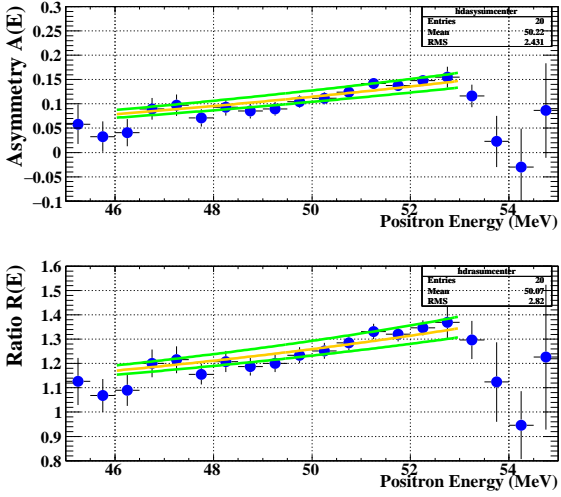}
\caption{\label{fig:fitresults} 
Fit of $A\left( E_{\rm e^{+}} \right)$ (upper plot) and 
of $R\left( E_{\rm e^{+}} \right)$ (lower plot) as a function of the 
positron energy. The experimental data 
are corrected, year by year, by the MC-based tracking efficiency functions 
and the fitting function for the $\langle \cos \theta \left( E \right)\rangle$ 
dependence. The green lines represent the $\pm 1~\sigma$ band, including both  
the statistic and the systematic uncertainties. 
The fitting functions are obtained from the distribution in Eq.~\ref{micheldist}.} 
\end{figure}

A quantitative estimate of the polarization can be obtained by studying 
the angle-integrated energy distributions $\left( dN/dE \right)_{US}$ and 
$\left( dN/dE \right)_{DS}$ on the US and DS sides. Eq.~\ref{micheldist} 
shows that the difference between the US and DS sides is 
due to the presence of a term proportional to $x P_{\mu} \cos \theta$. Since 
the sign of this 
term changes from US (where, according to our definition of polar angles, 
it is positive) to DS (where, with the same definition, it is negative), 
one expects that both the asymmetry $A\left( E_{\rm e^{+}} \right)$ and the 
ratio $R\left( E_{\rm e^{+}} \right)$ increase almost linearly with the 
positron energy. The slope of this dependence is $P_{\mu} \cos \theta$: 
one can therefore extract a polarization value by fitting the experimentally 
measured asymmetry and ratio and dividing the measured slope by the average 
value of $\langle \left| \cos \theta \right| \rangle = 0.1762$ for the US and DS sections. 
However, since the angular acceptance is correlated with the energy, 
the averarge value of $\cos \theta$ is a function $\langle \cos \theta \left( E \right)\rangle$, 
which can be extracted directly from the data. Then, we replaced in the fitting 
formula the energy averaged value $\langle \left| \cos \theta \right| \rangle$ 
with $\langle \cos \theta \left( E \right)\rangle$, bin by bin (the differences  
between the energy dependent values and the energy averaged one are at 
$\pm 5 \,\%$ level). The fit interval was restricted to $\left( 46 - 53 \right)~{\rm MeV}$ 
to minimize possible distorsions 
due to the energy-angle dependencies of the energy threshold and because 
Eq.~\ref{micheldist} is meaningless for $E > 52.83~{\rm MeV}$, i.e. 
$x > 1$. The expected plots for $A\left( E_{\rm e^{+}} \right)$ and 
$R\left( E_{\rm e^{+}} \right)$ are shown in Fig.~\ref{fig:michelasyrat}. 
The experimental $A\left( E_{\rm e^{+}} \right)$ and 
$R\left( E_{\rm e^{+}} \right)$ were separately determined year by year 
and summed. The fit results for the full data sample are shown in 
Fig.~\ref{fig:fitresults}; the average value of the two fits is 
\begin{equation}
P_{\mu} = -0.856 \pm 0.021,
\label{eq:fitresults}
\end{equation} 
where the quoted error is only statistical. 
The average $\chi^{2}/d.o.f.$ of the fits is $0.74$, mainly determined by the 
points close to the threshold. In both plots the yellow line represents the 
best fit, while the two green lines show the $\pm 1~\sigma$ band, obtained 
by adding or subtracting the sum of statistic and systematic uncertainties 
(see next section for the discussion of systematic uncertainties). 

If we fit the polarization values year by year we obtain the results 
reported in Tab.~\ref{tab:polfitbyyear}, where again the quoted errors 
are only statistical. 
\begin{table}
\caption{\label{tab:polfitbyyear} Results of the polarization fit 
year by year.}
\begin{tabular*}{\columnwidth}{@{\extracolsep{\fill}}lcccc@{}}
\hline
\hline
                                 & $ 2009 $         & $ 2010 $         & $ 2011 $         & Global \\
\hline
$\langle P \rangle \pm \Delta P$ & $ 0.85 \pm 0.05$ & $ 0.71 \pm 0.05$ & $ 0.92 \pm 0.04$ & $ 0.856 \pm 0.021 $ \\
$\chi^{2}$/d.o.f.                & $ 0.90 $         & $ 1.37 $         & $ 0.34 $         & $ 0.74 $        \\
\hline
\hline
\end{tabular*}
\end{table}
The polarization value measured in $2010$ sample is significantly lower than in the 
other two years; however, the larger values of the $\chi^{2}$/d.o.f. suggests that 
this result is of lower quality and less reliable. 
The observed deviation in $2010$ data is discussed in the next section 
and is reflected in the associated systematic error. 

\begin{figure}[hbt]
\includegraphics[width=0.5\textwidth]{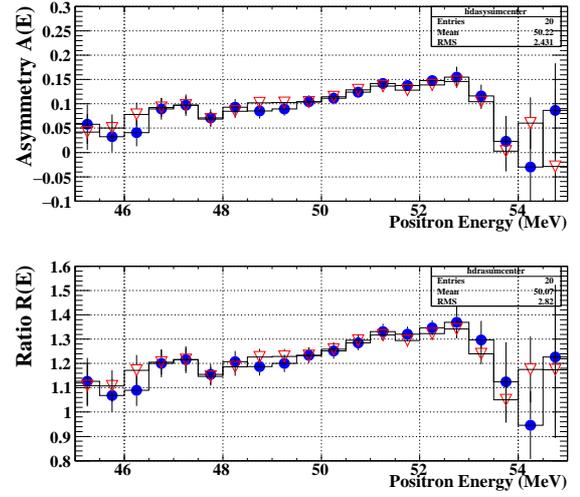}
\caption{\label{fig:compamc085}
Comparison between data (blue filled points) and MC for $P_{\mu} = -0.85$ 
(red open triangles). $A\left( E_{\rm e^{+}} \right)$ (upper plot) and 
$R\left( E_{\rm e^{+}} \right)$ 
(lower plot) between $45$ and $55~{\rm MeV}$.} 
\end{figure}

In Fig.~\ref{fig:compamc085} we show the 
comparison between data (blue filled points) and MC generated with 
$P_{\mu} = -0.85$ (red open triangles) for $A\left( E_{\rm e^{+}} \right)$ 
(upper plot) and $R\left( E_{\rm e^{+}} \right)$ 
(lower plot) between $45$ and $55~{\rm MeV}$: the agreement is quite good 
everywhere in the selected energy interval. 
Note that the energy region above $53~{\rm MeV}$, where the 
data and MC errors are quite large, does not affect the result, 
since the fit was limited to $\left( 46 - 53 \right)~{\rm MeV}$. 
We checked that these results do not depend on the fitting interval by 
eliminating one bin at the lower bound and/or one bin at the upper bound: 
in all cases the fit results agreed with (\ref{eq:fitresults}) within the 
statistical error. 
\subsubsection{Systematic uncertainties}\label{sec:syst}
Various systematic uncertainties can produce sizable effects on this measurement. 
We single out seven main possible sources: energy scale, 
angular bias, target position, MC-based efficiency 
corrections, threshold effects, higher order corrections, 
including the effect of finite electron mass, in the theoretical 
calculations and off-target muon decays. 
The first three affect 
the shape of the spectra on the US and DS sides and the evaluation of 
the relative efficiency $\epsilon \left( TC \left| Track \right. \right)$ 
from data; the fourth determines the absolute tracking 
efficiency; the fifth can alter the $A\left( E_{\rm e^{+}} \right)$ and 
$R\left( E_{\rm e^{+}} \right)$ fits in the bins close to the 
lower bound of the fit interval, the sixth can modifiy the 
fitting function and the seventh can alter the quality of the selected 
positron sample.
\begin{itemize}
\item [1)] Energy scale. The energy scale and resolution are 
determined in MEG, as discussed in \cite{meg2010}, by fitting the Michel 
positron energy spectrum with the convolution of the theoretical 
spectrum (including radiative corrections) of the detector 
acceptance and of a resolution curve, in the form of a partially 
constrained triple Gaussian shape. The position of the Michel edge, 
used as a reference calibration point, is determined with a precision of 
$\delta E_{\rm e^{+}} \sim 30~~{\rm keV}$. The effect of this uncertainty 
was evaluated by varying the reconstructed energy of our events by a factor 
$\left( 1 \pm \delta E_{\rm e^{+}}/\bar{E}_{\rm e^{+}} \right)$, where 
$\bar{E}_{\rm e^{+}} = 52.83~{\rm MeV}$ is the position of the Michel edge, and 
repeating the analysis. The polarization value determined by 
the average of $A\left( E_{\rm e^{+}} \right)$ and 
$R\left( E_{\rm e^{+}} \right)$ fits increases by $0.0029$ when 
$\delta E_{\rm e^{+}}/\bar{E}_{\rm e^{+}}$ is added and decreases 
by $-0.0052$ when $\delta E_{\rm e^{+}}/\bar{E}_{\rm e^{+}}$ is subtracted.
\item [2)] Angular bias. The angular resolution is determined 
by looking at tracks crossing the chamber system twice (double 
turn method), as discussed in \cite{meg2009,meg2010}. 
The uncertainty on the $\theta$ and $\phi$ scales varies between 
$1$ and $3~{\rm mrad}$. The effect of this uncertainty on 
the angular scale was (conservatively) evaluated by modifying 
both the reconstructed polar angles by $\pm 3~{\rm mrad}$ and 
repeating the analysis. The measured polarization decreases 
(increases) by $-0.013$ ($+0.025$). 
\item [3)] Target position. The target position with respect 
to the centre of the COBRA magnet is measured by means of 
an optical survey and checked by looking at the distribution 
of the positron vertex of reconstructed tracks. The discrepancies 
between the two methods are at the 
level of a fraction of a mm. Since in our analysis we require that the 
reconstructed positron vertex lies within the target ellipse, an 
error on the target position can alter the positron selection. 
We assumed a conservative estimate of a target position uncertainty of 
$\pm 1~{\rm mm}$ on all coordinates 
and, as previously, added or subtracted it and repeated the analysis. 
The effect was to decrease (increase) the polarization by $-0.022$ 
($0.016$). 
\item [4)] MC-based corrections. The MC corrections, inserted to take 
into account the absolute tracking efficiency, are ba\-sed on the position 
of the target as measured by the optical survey and on the nominal 
location of the beam centre. A variation 
of these parameters produces a variation on the correction functions, 
applied year by year to MC and data. We estimated the size of 
this effect by generating MC samples with a displaced beam and target 
($\pm 1~{\rm mm}$ shift as previously) and null 
polarization and determined new tracking efficiency correction 
functions. Such functions were then applied to the data and MC: the 
measured polarization decreased (increased) by $-0.035$ ($0.036$). 
\item [5)] Threshold effects. The response of the MEG tracking 
system close to the momentum threshold ($E_{\rm e^{+}} \approx 
45~{\rm MeV}$) depends in general on the polar angles and can 
be significantly distorted when the beam and target are not centred, 
causing fictitious differences between the US and DS sides. 
In 2010 the beam centre to target centre displacement was maximal, 
corresponding to more than $3~{\rm mm}$ in the horizontal plane 
and just over $3~{\rm mm}$ in the vertical plane,
producing an asymmetric US-DS energy threshold,
with the DS spectrum systematically higher than the
US one for $E_{\rm e^{+}} < 47~{\rm MeV}$. 
The beam and target displacement were introduced in the MC, 
but the simulation for $2010$ did not result in a good agreement with 
the data in the region close to the energy threshold. We then estimated the 
systematic effect due to the angular dependence of the energy 
threshold by removing the $2010$ sample from the fit: 
the polarization decreases by $-0.047$, a difference twice larger than 
the statistical error. The $\chi^{2}/d.o.f.$ of the fit improved a 
bit from $0.74$ to $0.71$. A better fit quality was observed also on 
MC events by removing the simulated data corresponding to the 
year $2010$ configuration. 
\item [6)] Higher order corrections to the theoretical formula in Eq.~\ref{micheldist}. 
The effect of second and higher order contributions and of taking into account 
the finite electron mass to the muon decay rate is discussed
in some detail in \cite{Arbuzov:2002cn,Arbuzov:2002pp,Arbuzov:2002rp,Arbuzov:2004wr}.
The conclusion is that they are smaller than the first order correction and therefore
we can deduce that the effect of including them in Eq.~\ref{micheldist}
for extracting the polarization from Fig.~\ref{fig:fitresults} is not larger than
the effect of the first order correction that is 0.3\%. Hence this value can be 
assumed as a conservative estimation of the systematic error due to higher order corrections.
\item [7)] Off-target muon decays. A conservative estimation of off-target muon decays as discussed in 
Sect.~\ref{sect:analysis} is 0.5\%, that is 0.004 on the polarization value. 
\end{itemize}
The effects of the various systematic uncertainties and the global 
systematic uncertainty calculated by their addition in quadrature are 
reported in Tab.\ref{tab:syst}.

\begin{table}
\caption{\label{tab:syst} Main systematic uncertainties and their 
effect on polarization.}
\begin{tabular*}{\columnwidth}{@{\extracolsep{\fill}}lc@{}}
\hline\hline
Source & ($\Delta P$) \\\hline
Energy scale & $\left( +0.0029, -0.0052 \right)$ \\
Angular scale & $\left( +0.025, -0.013 \right)$ \\
Target position & $\left( +0.016, -0.022 \right)$ \\
Tracking efficiency & $\left( +0.036, -0.035 \right)$ \\
Energy threshold & $-0.047$ \\
Higher order corrections & $\pm 0.003$ \\
Off-target decsy & $\pm 0.004$ \\
\hline
Total (in quadrature) & $\left( +0.047, -0.064 \right)$ \\
\hline\hline
\end{tabular*}
\end{table}
Combining the fit results in (\ref{eq:fitresults}) with the numbers 
reported in Tab.\ref{tab:syst} we can state that the muon residual 
polarization in the MEG experiment is:
\begin{equation}
P_{\mu} = -0.86 \pm 0.02 ~ {\rm (stat)} ~ { }^{+ 0.05}_{-0.06} ~ {\rm (syst)}.
\label{eq:polmeas}
\end{equation} 
\section{Summary and conclusions}
We measured the residual muon polarization $P_{\mu}$ in the 
MEG experiment 
by studying the energy-angle distribution 
of Michel positrons collected during three years of data 
taking. We obtained:
\begin{equation}
P_{\mu} = -0.86 \pm 0.02 ~ {\rm (stat)} ~ { }^{+ 0.05}_{-0.06} ~ {\rm (syst)}.
\label{eq:summary}
\end{equation} 
The measured value is in agreement with the value expected from calculation of 
the depolarizing effects due to the muon spin interactions during the production 
and the propagation through the apparatus up to the stopping target, based 
on the SM prediction of positive surface muons, produced fully polarized 
in the direction opposite to the beam direction. Moreover, 
the Michel positron angular distribution 
and the US - DS asymmetry of the positron energy spectra are well 
reproduced by a complete simulation of the positron detection in the MEG set-up 
when a muon polarization $P_{\mu} = -0.85$ is used as an input parameter in 
the MC calculation. 
This result is important to allow a precise calculation of the Radiative 
Muon Decay branching ratio and energy-angle distribution in the 
kinematic region where it represents a background source to the search 
for ${\megsign}$ and can be used as a tool for the absolute normalization 
of the MEG experiment.   
\section{Acknowledgements}
We are grateful for the support and cooperation provided by PSI as 
the host laboratory and to the technical and engineering staff of our 
institutes. This work is supported by SNF grant 200021\_137738 (CH), DOE 
DEFG02-91ER40679 (USA), INFN (Italy) and MEXT KAKENHI 22000004 and 
26000004 (Japan). 
Partial support of the Italian Ministry of University and Research 
(MIUR) grant RBFR08XWGN, Ministry of University and Education of 
the Russian Federation and Russian Fund for Basic Research grants 
RFBR 14-22-03071 are acknowledged. 

\bibliographystyle{unsrt}
\bibliography{MEG}

\end{document}